\RequirePackage{fix-cm}
\documentclass[twocolumn]{svjour3}          
\smartqed  
\usepackage{graphicx}
\usepackage[latin1]{inputenc}
\usepackage{latexsym}
\usepackage{amssymb}
\usepackage{float}

\begin{document}

\title{Clifford Algebra and Space-Time Transformations 
}
\subtitle{Lorentz Transformation and Inertial Transformation}


\author{R. de Oliveira        \and
        S. J. da Silva \and V. H. G. de Campos
}


\institute{R. de Oliveira, \email{rosevaldo@ufmt.br} \and S. J. da Silva \email{oluasjose66@hotmail.com} \and V. H. G. de Campos \email{vichampos@hotmail.com}
                \at
                Departamento de Matemática, Universidade Federal do Mato Grosso - UFMT  \\
                Tel.: +55-66-34104028}

\date{Received: November 25th, 2013.}

\maketitle

\begin{abstract}
We review the Inertial transformation and Lorentz transformation under a new context, by using Clifford Algebra or Geometric Algebra. The apparent contradiction between theses two approach is simply stems from different procedures for clock synchronization associated with different choices of the coordinates used to describe the physical world. We find the physical and coordinates components of both transformations. A important result is that in the case of Inertial transformation the physical components are exactly the Galilean transformations, but the speed of light is not $c$. Another interesting result is due to the fact the Lorentz transformations lead directly to physical components, and this case the speed of light is $c$. Finally e show that both scenarios, de-synchronization Einstein theory and synchronized theory, are all mathematically equivalent by means of Clifford Algebra Transformations.

\end{abstract}

\section{Introduction}

In Einstein's special theory of relativity \cite{Einstein}, the Lorentz transformations are derived by postulating the relativity principle and the constancy of the speed of light. Nowadays, physicist agree that the Lorentz Transformation describes a fundamental symmetry of all natural phenomena. However, it would be interesting to know if there is any alternative to such transformation. We can begin by citing the transformation of Robertson \cite{Robertson} as one of the first attempts in this direction, Robertson proposed to replace the Einstein'postulates with hypotheses suggested by certain typical optical experiments as a way of testing the Einstein relativity theory. The most interesting member of this family of transformation was found by Tangherlini \cite{Tangherlini1,Tangherlini2} and studied by Mansouri and Sexl \cite{Mansouri1,Mansouri2,Mansouri3}, Marinov \cite{Marinov}, Chang \cite{Chang} and Rembieliski \cite{Rembielinski} among others. According to Lorentz-Poincaré; Reinchenbach; Mansouri and Sexl and Jammer the clock synchronization in inertial systems is conventional, Einstein explicitly agreed in considering this part of his theory conventional. However, Franco Selleri \cite{Selleri} don't agre. Selleri argue that there exist experimental facts against the Einstein Synchronization choice, he ensures that it is not possible to explain the Sagnac effect by using Einstein synchronization procedure. Selleri goes beyond this issue, he argues that the only transformations which describe the experimental data are the Inertial Transformations or Synchronized transformation. In this work we are not interested in proving what is the correct method of synchronizing clocks, or to find out who is right in this dispute. We will show that using Clifford Algebra we can obtain both transformations, Lorentz Transformation and Inertial Transformation, by a same mathematically consistent way.

\section{Clifford Algebra of Space-Time $\mathbb{R}^{2}$}
For simplicity we consider a two-dimensional vectorial space over
the field of reals, $\mathbb{R}^{2}$. The generalization to higher
dimensions has no complications. The canonical bases are defined as
$(\hat{e}_{0},\hat{e}_{1})$ in this space. A vector in this space
assume the following form
\begin{eqnarray}
\vec{R}=x^{\mu}\hat{e}_{\mu}=x^{0}\hat{e}_{0}+x^{1}\hat{e}_{1}
\end{eqnarray}
where $x^{0}=c t$ and $x^{1}=x$. With $c$ is the speed of light and
$(ct,x)$ is the time and space coordinate of an event. Until now we
did not define what kind of algebra we wish to adopt. We will stand
for this purpose the following condition under our vectors
\begin{eqnarray}
|\vec{R}|^{2}=\vec{R}\vec{R}=(x^{0})^{2}-(x^{1})^{2}
\end{eqnarray}
We can clearly see that the Pythagorean theorem is not valid in this
space. The unit base vectors need to obey the following algebra
\begin{eqnarray}
\hat{e}_{0}^{2}=1\qquad \hat{e}_{1}^{2}=-1\\
\hat{e}_{0}\hat{e}_{1}+\hat{e}_{1}\hat{e}_{0}=0,
\end{eqnarray}
this algebra is known as Clifford algebra in 1+1 dimensions $({\cal
C}l_{1,1})$ \cite{h1,h2,h3,h4,h5,h6,h7,h8,h9}. Given two vectors $a$ and $b$ belonging to this space,
the product of symmetric and anti symmetric algebra are given by
\begin{eqnarray}
a\cdot b&=&\frac{1}{2}(ab+ba)\,\,{\it symmetric} \\
a\Lambda b&=&\frac{1}{2}(ab-ba)\,\,{\it
anti symmetric}
\end{eqnarray}

The metric tensor in this space is defined by the symmetric product as shown below
\begin{eqnarray}
g_{\mu\nu}=\hat{e}_{\mu}\cdot\hat{e}_{\nu}=\frac{1}{2}(\hat{e}_{\mu}\hat{e}_{\nu}+\hat{e}_{\nu}\hat{e}_{\mu})
\end{eqnarray}
\begin{eqnarray}
[g_{\mu\nu}]=\left(\begin{array}{cc} 1 & 0 \\ 0 & -1\end{array}\right)
\end{eqnarray}

The covariant components are written using the following relation
\begin{eqnarray}
x_{\mu}&=&\vec{R}\cdot\hat{e}_{\mu}=\frac{1}{2}(\vec{R}\hat{e}_{\mu}+\hat{e}_{\mu}\vec{R})\\
x_{\mu}&=&g_{\mu\nu}x^{\nu}
\end{eqnarray}
The contravariant bases are defined by using the contravariant
tensor metric
\begin{eqnarray}
\hat{e}^{\mu}=g^{\mu\nu}\hat{e}_{\nu}
\end{eqnarray}
where $g^{\mu\nu}$ is defined to be the inverse of $g_{\mu\nu}$, as
shown below
\begin{eqnarray}
g^{\mu\rho}g_{\rho\nu}=\delta^{\mu}_{\nu}
\end{eqnarray}
where $\delta^{\mu}_{\nu}$ is the well known Kronecker delta. In a
different way, we can write the contravariant basis vectors in
matrix representation
\begin{eqnarray}
\left(\begin{array}{c}\hat{e}^{0}\\ \hat{e}^{1}\end{array}\right)=
\left(\begin{array}{cc} 1 & 0\\ 0 & -1\end{array}\right)
\left(\begin{array}{c} \hat{e}_{0} \\ \hat{e}_{1}\end{array}\right)
\end{eqnarray}
where the contravariant tensor metric is given by
\begin{eqnarray}
[g^{\mu\nu}]=\left(\begin{array}{cc} 1 & 0\\ 0 &
-1\end{array}\right)
\end{eqnarray}

\section{Clifford Algebra and Passive Transformation}
In this section we are interested in finding passive transformations
between two inertial reference systems. These changes preserve the
vector $\vec{R'}=\vec{R}$ and the module of the original vector
$|\vec{R'}|=|\vec{R}|$. The transformation must obey the following
relations
\begin{eqnarray}
\vec{R'}&=&x'^{0}\hat{e}'_{0}+x'^{1}\hat{e}'_{1}=\vec{R}=x^{0}\hat{e}_{0}+x^{1}\hat{e}_{1}\\
\vec{R'}&=&x'^{\mu}\hat{e}'_{\mu}=x^{\mu}\hat{e}_{\mu}
\end{eqnarray}
We define the transformation of coordinates and bases in the
following way
\begin{eqnarray}
x'^{\mu}=\Lambda^{\mu\nu}x_{\nu}\qquad {\it and} \qquad
\hat{e}'_{\mu}=\Gamma_{\mu\nu}\hat{e}^{\nu}
\end{eqnarray}
The transformation of the vector $\vec{R}$ gives us
\begin{eqnarray}
\vec{R}=x'^{\mu}\hat{e}'_{\mu}=\Lambda^{\mu\nu}\Gamma_{\mu\rho}x_{\nu}\hat{e}^{\rho}
\end{eqnarray}
and to the vector be invariant the following relation need to be
satisfied
\begin{eqnarray}
\Lambda^{\mu\nu}\Lambda_{\mu\rho}=\delta^{\nu}_{\rho}
\end{eqnarray}
we have defined the inverse as $\Gamma_{\mu\nu}=\Lambda_{\mu\nu}$.

Therefore in two-dimensional space-time, theses transformation need
to obey the four conditions
\begin{eqnarray}
\Lambda^{00}\Lambda_{00}+\Lambda^{10}\Lambda_{10}=1 \qquad
\Lambda^{00}\Lambda_{01}+\Lambda^{10}\Lambda_{11}=0\\
\Lambda^{01}\Lambda_{00}+\Lambda^{11}\Lambda_{10}=0 \qquad
\Lambda^{10}\Lambda_{01}+\Lambda^{11}\Lambda_{11}=1
\end{eqnarray}
To solve this system of equations we must necessarily impose desired
physical conditions.

\begin{figure*}[tp] 
\centering
\includegraphics{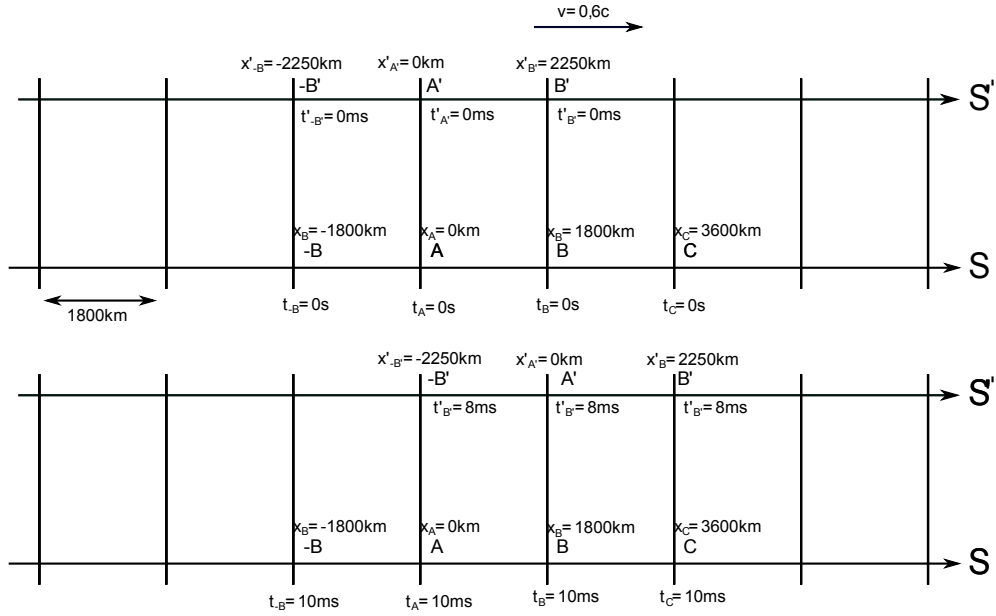}
\caption{Synchronized transformation}\label{fig1}
\label{fig:IT}
\end{figure*}

\subsection{Space Physical Condition}
The space condition appears to be quite intuitive. If an object
(point) is at the origin of the system S' $(x'^{1}= 0)$, then its
position in the system S $(x^{1})$ is given by $x^{1}=v t$. Where
$v$ is the velocity of the system S'. We are considering that $t=0$
when the origins of the systems coincide. A direct consequence of
this condition is given by
\begin{eqnarray}
x'^{1}&=&0 \to x^{1}=vt\qquad,\qquad 0=\Lambda^{10}x_{0}+\Lambda^{11}x_{1}\\
\frac{v}{c}&=&\frac{\Lambda^{10}}{\Lambda^{11}}
\end{eqnarray}

\subsection{Time Physical condition or Synchronization Choice}
  Let $S$ denote the rest system and $S'$ the moving one. The relative velocity of $S'$ to $S$ is $v=0,6c$ along the $x$-axis of $S$. We define that all clocks in $S$ are synchronized. When the clocks in $S$ are at $t = 0 s$, what is the time value $t'$ of a clock at $S'$?  To answer this question a further condition must be satisfied
\begin{eqnarray}
ct'=\Lambda^{00}ct-\Lambda^{01}x^{1} \to ct'=-\Lambda^{01}x^{1}
\end{eqnarray}
If we choose $\Lambda^{01}\neq 0$, all the clocks in the system $S'$ will be de-synchronized. Otherwise if we choose $\Lambda^{01}=0$ all the clocks in $S'$ will be synchronized. We should now make a choice that is a matter of convention.

\subsubsection{Synchronized Choice}

Consider that there exist a system $S$ that the one-way speed of light in empty space is $c$ in any direction. We can synchronize the clocks of the system $S$ with the usual Einstein procedure involving light rays \cite{Abreu}, since the one-way speed of light in this system is known. Another system $S'$ moving with respect to system $S$ with velocity $v$. This system $S'$ has a set of clocks that need to be synchronized in some way. We will obtain the synchronized transformation if we consider that all clocks in this system $S'$ are marking the same time. If we synchronize all clock of system $S'$ in a way that all clocks mark the same time, the coordinates transformations will not be the well known Lorentz transformation. As a result we will obtain different transformations, called as Synchronized Transformation or Inertial transformation. They were obtained by Mansouri and Sexl in 1977 \cite{Mansouri1,Mansouri2} and have been emphasized by Franco Selleri \cite{Selleri}.

The figure-\ref{fig1} illustrate the behavior of the Synchronized Transformation. In this example we are using $v=0,6 c$, and $L=1800$ km

The phenomenon of time dilatation can be deduced in the usual way, such as presented in the Feynman'book \cite{Feynman}, using a light clock placed in $S'$ aligned along $y$-axis. The well result is
\begin{eqnarray}
\Delta t=\gamma \Delta t'
\end{eqnarray}
with
\begin{eqnarray}
\gamma=\frac{1}{\sqrt{1-\frac{v^{2}}{c^{2}}}}
\end{eqnarray}
express the fact that ``moving clocks run slower".

\begin{figure*}[tp]
\centering
\includegraphics{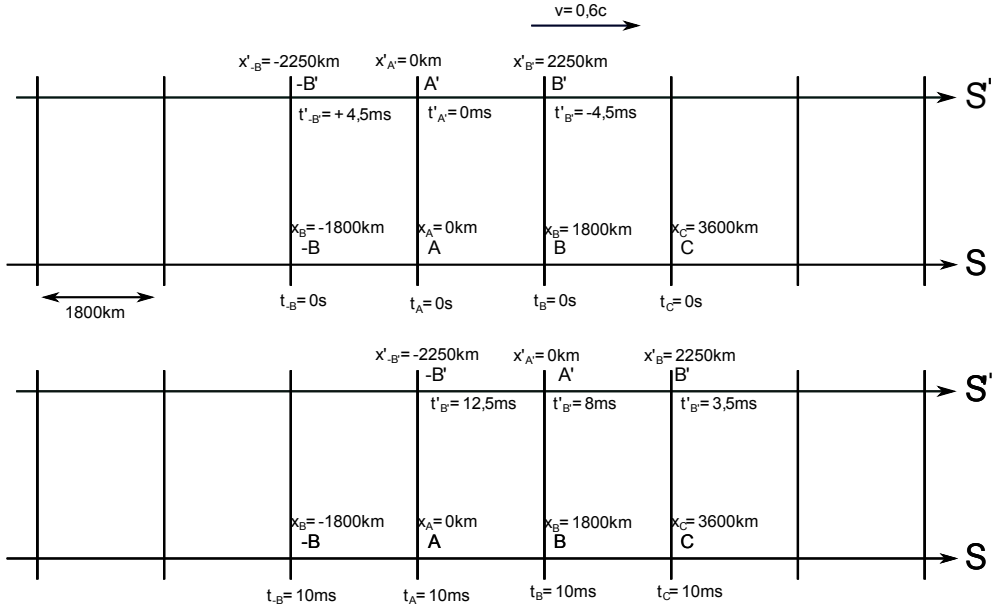}
\caption{De-synchronized Transformation}\label{De-synchronized}
\end{figure*}

The phenomenon of space contraction can be deduced as well in the usual way, using a light clock placed in $S'$ aligned the $x$-axis. The result is
\begin{eqnarray}
L'=\gamma L
\end{eqnarray}
it say that ``moving ruler are shorter".

The transformation of coordinates between $S$ and $S'$ can now be obtained
\begin{eqnarray}
x'&=&\gamma(x-vt)\\
t'&=&\frac{t}{\gamma}
\end{eqnarray}
the above transformation are known as Synchronized Transformation or Inertial Transformation.

Finally we can write the complete transformation in a matrix way
\begin{eqnarray}
\left(\begin{array}{c}
x'^{0}\\ x'^{1}\end{array}\right)&=&\left(\begin{array}{cc}
{\gamma}^{-1} & 0\\ -\frac{\gamma v}{c} & -\gamma \end{array}\right)\left(\begin{array}{c}
x_{0}\\ x_{1}\end{array}\right)
\end{eqnarray}

\begin{eqnarray}
\left(\begin{array}{c}
\hat{e}'_{0}\\ \hat{e}'_{1}\end{array}\right)&=&\left(\begin{array}{cc}
\gamma &- \frac{\gamma v}{c}\\0&-{\gamma}^{-1}\end{array}\right)\left(\begin{array}{c}
\hat{e}^{0}\\ \hat{e}^{1}\end{array}\right)
\end{eqnarray}

\begin{figure}[H]
\begin{center}
\includegraphics{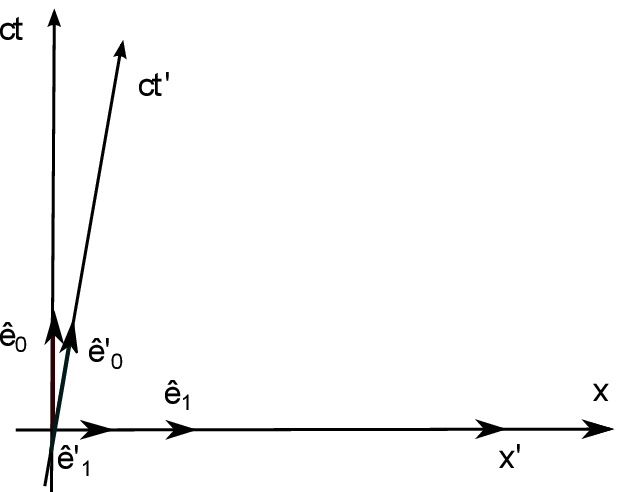}
\caption{Basis transformation}\label{fig2}
\end{center}
\end{figure}

We represent the basis transformation in the figure-\ref{fig2}

We write explicitly the transformations of the basis below
\begin{eqnarray}
\hat{e}'_{0}&=&\gamma\hat{e}^{0}-\frac{\gamma v}{c}\hat{e}^{1}\\
\hat{e}'_{1}&=&-\frac{1}{\gamma} \hat{e}^{1}
\end{eqnarray}

So the metric system in S 'is given by
\begin{eqnarray}
g'_{\mu\nu}=\left(\begin{array}{cc}
1 & -{v}/{c}\\ -{v}/{c} & -\frac{1}{\gamma^{2}}\end{array}\right)
\end{eqnarray}

\begin{eqnarray}
g'^{\mu\nu}=\left(\begin{array}{cc}
\frac{1}{\gamma^{2}} & -{v}/{c}\\ -{v}/{c} & -1\end{array}\right)
\end{eqnarray}

We can verify that the speed of light in $S'$ is given by
\begin{eqnarray}
\Delta s^{2}=d\vec{R}\,d\vec{R}=g'_{\mu\nu}dx'^{\mu}dx'^{\nu}=0\\
V'=\frac{dx'}{dt'}=\gamma^{2}(-v\pm c)
\end{eqnarray}

\subsubsection{De-synchronized Choice - or Einstein Choice}
The Einstein Synchronization choice is the one that $\Lambda^{01}\neq 0$.
The Lorentz transformations are written as follows
\begin{eqnarray}
x''^{0}&=&\gamma (x_{0}+\beta x_{1})\\
x''^{1}&=&\gamma (-x_{1}-\beta x_{0})
\end{eqnarray}
where $x^{0}=c t$, $x''^{0}=c t''$, $\beta=\frac{v}{c}$ and
$\frac{1}{\gamma^{2}}=1-\beta^{2}$. A different way to write these
transformations is through the matrix representation
\begin{eqnarray}
\left(\begin{array}{c} x''^{0}\\
x''^{1}\end{array}\right)=\left(\begin{array}{cc} \gamma &
\beta\gamma\\-\beta\gamma & -\gamma\end{array}\right)\left(\begin{array}{c} x_{0}\\
x_{1}\end{array}\right)
\end{eqnarray}

\begin{eqnarray}
\left(\begin{array}{c} x_{0}\\
x_{1}\end{array}\right)=\left(\begin{array}{cc} \gamma &
\beta\gamma\\-\beta\gamma & -\gamma\end{array}\right)\left(\begin{array}{c} x''^{0}\\
x''^{1}\end{array}\right)
\end{eqnarray}
its important to note that this transformation is symmetric.


The figure-\ref{De-synchronized} illustrate the behavior of the \linebreak de-synchronized Transformation. In this example we are using $v=0,6 c$, and $L=1800$ km

And the basis are defined in the following way
\begin{eqnarray}
\hat{e}''_{0}&=&\gamma\hat{e}^{0}-\beta\gamma\hat{e}^{1}\\
\hat{e}''_{1}&=&\beta\gamma \hat{e}^{0}-\gamma \hat{e}^{1}
\end{eqnarray}

\begin{figure}[H] 
 \centering
 \includegraphics{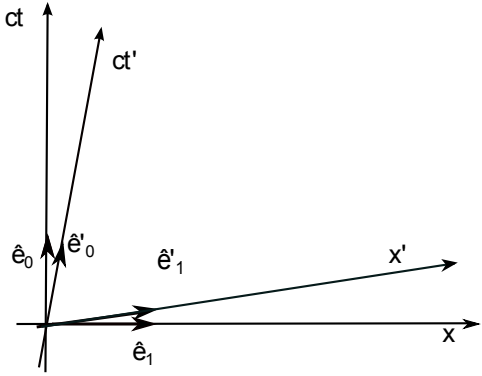}
 \caption{Basis transformation}\label{Einstein}
 \label{Lorentz Transformation}
 \end{figure}

The figure-\ref{Einstein} express theses transformation graphically.

In a matrix representation we can write the basis as follows
\begin{eqnarray}
\left(\begin{array}{c}
\hat{e}''_{0}\\ \hat{e}''_{1}\end{array}\right)&=&\left(\begin{array}{cc}
\gamma &-\gamma\beta\\ \gamma \beta &-{\gamma}\end{array}\right)\left(\begin{array}{c}
\hat{e}^{0}\\ \hat{e}^{1}\end{array}\right)
\end{eqnarray}

So the metric system in $S'$ is given by
\begin{eqnarray}
g''_{\mu\nu}=g''^{\mu\nu}=\left(\begin{array}{cc}
1 & 0\\ 0 & -1\end{array}\right)
\end{eqnarray}

note that the metric in the system $S'$ is the same one the system $S$.

It is easy to check that the speed of light in $S'$ is given by
\begin{eqnarray}
\Delta s^{2}=d\vec{R}\,d\vec{R}=g''_{\mu\nu}dx''^{\mu}dx''^{\nu}=0\\
V''=\frac{dx''}{dt''}=c
\end{eqnarray}

\section{Transformation between Synchronized and Desynchronized Choices}

Let us introduce the following notation in matrix representation as follows
\begin{eqnarray}
\vec{x}_{c}=\left(\begin{array}{c} x_{0}\\
x_{1}\end{array}\right),\qquad \vec{x}^{c}=\left(\begin{array}{c} x^{0}\\
x^{1}\end{array}\right)\\
\qquad [g]_{c}=\left(\begin{array}{cc} g_{00} &
g_{01}\\g_{10} & g_{11}\end{array}\right)\\
\qquad [g]^{c}=\left(\begin{array}{cc} g^{00} &
g^{01}\\g^{10} & g^{11}\end{array}\right)
\end{eqnarray}
\begin{eqnarray}
\vec{e}_{c}=\left(\begin{array}{c} \hat{e}_{0}\\
\hat{e}_{1}\end{array}\right),\qquad \vec{e}^{c}=\left(\begin{array}{c} \hat{e}^{0}\\
\hat{e}^{1}\end{array}\right)
\end{eqnarray}

\begin{eqnarray}
[g]_{c}\cdot [g]^{c}=1
\end{eqnarray}
\begin{eqnarray}
\vec{x}_{c}=[g]_{c}\vec{x}^{c} \quad \vec{x}^{c}=[g]^{c}\vec{x}_{c}
\end{eqnarray}
The transformation between two systems is given by
\begin{eqnarray}
\vec{x'}^{c}=\mathbb{M}^{t}\vec{x''}_{c}
\end{eqnarray}
\begin{eqnarray}
[g'']_{c}^{-1}=\mathbb{M}[g']_{c}\mathbb{M}^{t}
\end{eqnarray}

If you do not care about the problem with the physical components, we can write the transformation between
synchronized and desynchronized choices by means of the following matrix transformation
\begin{eqnarray}
\mathbb{M}^{t}=\left(\begin{array}{cc} 1 &
-\beta\\0 & -1\end{array}\right)
\end{eqnarray}

Explicitly the coordinates transform as
\begin{eqnarray}
\left(\begin{array}{c} x'^{0}\\
x'^{1}\end{array}\right)=\left(\begin{array}{cc} 1 &
-\beta\\0 & -1\end{array}\right)\left(\begin{array}{c} x''_{0}\\
x''_{1}\end{array}\right)
\end{eqnarray}

and the basis transform as follows
\begin{eqnarray}
\left(\begin{array}{c} \hat{e}'_{0}\\
\hat{e}'_{1}\end{array}\right)=\left(\begin{array}{cc} 1 &
0\\-\beta & -1\end{array}\right)\left(\begin{array}{c} \hat{e}''^{0}\\
\hat{e}''^{1}\end{array}\right)
\end{eqnarray}

\section{Physical Components versus Coordinate Value}

\begin{figure}[H] 
 \centering
 \includegraphics[scale=0.5]{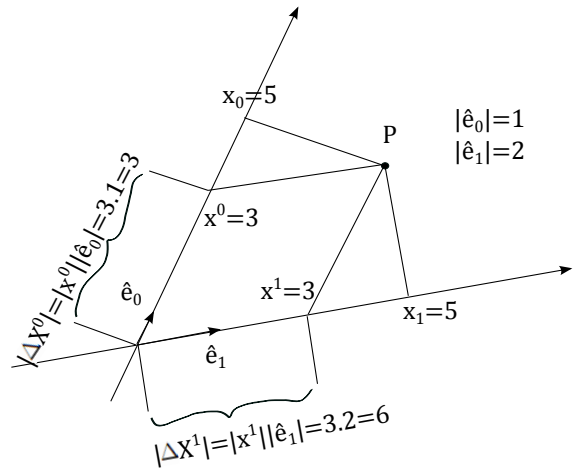}
 \caption{Non-orthogonal transformation} \label{non orthogonal}
 \end{figure}

For generalized coordinate systems, the numerical values of vectors components do not generally be the physical values. The physical values are measured with standard physical instruments. The figure-\ref{non orthogonal} illustrates the situation where a non-orthogonal coordinate grids do not preserves the basis as unit vectors. If we look at one component of a vector, it has a coordinate value and a physically measured value \cite{Klauber,Klauber2,Wrede}.

The physical vector distance $d\vec{R}$ between two points in two-dimensional vectorial space over
the field of reals $\mathbb{R}^{2}$ is defined by
\begin{eqnarray}
d\vec{R}=dX^{\mu}\hat{e}^{\ast}_{\mu}=dX^{0}\hat{e}^{\ast}_{0}+dX^{1}\hat{e}^{\ast}_{1}
\end{eqnarray}
where $dX^{\mu}$ are physical distances and $\hat{e}^{\ast}_{\mu}$ are unit basis vectors.

For the same vector $d\vec{R}$ expressed in a different coordinate system we can have different component coordinate $d\vec{X}^{\mu}\neq d\vec{x}^{\mu}$ and different basis vectors $\hat{e}_{\mu}\neq \hat{e}^{\ast}_{\mu}$, but the similar expression for distance
\begin{eqnarray}
d\vec{R}=dx^{\mu}\hat{e}_{\mu}=dx^{0}\hat{e}_{0}+dx^{1}\hat{e}_{1}
\end{eqnarray}
under a passive transformation
\begin{eqnarray}
dx^{\mu}&=&\Lambda^{\mu\nu}dX_{\nu}\\
\hat{e}_{\mu}&=&\Lambda_{\mu\sigma}\hat{e^{\ast}}^\sigma
\end{eqnarray}
where
\begin{eqnarray}
\Lambda^{\mu\nu}\Lambda_{\mu\sigma}=\delta^{\nu}_{\sigma}
\end{eqnarray}

Physical components are designates by unit vectors. Therefore if we are interested in finding the value of
physical components related to measured results we must take care with it.

We can find the physical values of the coordinate in the new system by
\begin{eqnarray}
d\vec{R}&=&dx^{\mu}\hat{e}_{\mu}=dx^{0}\alpha \frac{\hat{e}_{0}}{\alpha}+dx^{1}\beta \frac{\hat{e}_{1}}{\beta}\\
d\vec{R}&=&dx^{0}\alpha\hat{e}^{\ast}_{0}+dx^{1}\beta\hat{e}^{\ast}_{1}\\
d\vec{R}&=&dX^{0}\hat{e}^{\ast}_{0}+dX^{1}\hat{e}^{\ast}_{1}
\end{eqnarray}

So, the relation between physical components and coordinate components (mathematical entities) are
\begin{eqnarray}
dX^{0}=\alpha dx^{0}&\,\,\,\,\,&dX^{1}=\beta dx^{1}\\
\hat{e}^{\ast}_{0}=\frac{\hat{e}_{0}}{\alpha}&\,\,\,\,\,&\hat{e}^{\ast}_{1}=\frac{\hat{e}_{1}}{\beta}
\end{eqnarray}
For a detailed discussion of this subject see the work of Klauber \cite{Klauber2}.

\subsection{Physical Components and Synchronized Transformation}

The synchronized transformation or inertial transformation are given by
\begin{eqnarray}
\hat{e}'_{0}&=&\gamma\hat{e}_{0}+\frac{\gamma v}{c}\hat{e}_{1}\\
\hat{e}'_{1}&=&\frac{1}{\gamma}\hat{e}_{1}
\end{eqnarray}
and the metric of this system in $S'$ is
\begin{eqnarray}
g'_{\mu\nu}=\left(\begin{array}{cc}
1 & -\frac{v}{c}\\ -\frac{v}{c} & \frac{v^{2}}{c^{2}}-1\end{array}\right)
\end{eqnarray}
In order to obtain the physical components in the new reference system we choice a new unit basis as
\begin{eqnarray}
\hat{e}'^{\ast}_{1}=\gamma\hat{e}'_{1}
\end{eqnarray}

However the coordinates need to transform as following
\begin{eqnarray}
X'^{1}=\frac{x'^{1}}{\gamma}=\frac{\gamma(x-vt)}{\gamma}=x-vt
\end{eqnarray}
the physical value of spacial coordinate are like the Galilean transformation.

As the sacalar product $\hat{e}'_{0}\cdot \hat{e}'_{0}=1$, the physical value of the time coordinate do not change
\begin{eqnarray}
X'^{0}=x'^{0}\qquad {\it or}\qquad  T'=t'=\frac{t}{\gamma}
\end{eqnarray}

And the new metric tensor is
\begin{eqnarray}
g'^{\ast}_{\mu\nu}=\left(\begin{array}{cc}
1 & -\gamma\frac{v}{c}\\ -\gamma\frac{v}{c} & -1\end{array}\right)=\gamma^{2}(g'^{\ast})^{\mu\nu}
\end{eqnarray}

We can compute and verify that the distance is preserved
\begin{eqnarray}
\vec{R}\vec{R}&=&|\vec{R}|^{2}=(X'^{0})^{2}-(X'^{1})^{2}-2\beta\gamma X'^{0}X'^{1}\\
 &=&(x^{0})^{2}-(x^{1})^{2}
 \end{eqnarray}

The speed of light in $S'$ now is given by
\begin{eqnarray}\label{vnc}
\Delta s^{2}=d\vec{R}\,d\vec{R}=g'^{\ast}_{\mu\nu}dX'^{\mu}dX'^{\nu}=0\\
V'^{\ast}=\frac{dX'}{dT'}=\gamma(-v\pm c) \nonumber
\end{eqnarray}

\subsection{Physical Components and De-synchronized Transformation}
In this case the unit basis transform in the following way
\begin{eqnarray}
\hat{e}''_{0}&=&\gamma\hat{e}_{0}+\beta\gamma\hat{e}_{1}\\
\hat{e}''_{1}&=&\beta\gamma \hat{e}_{0}+\gamma \hat{e}_{1}
\end{eqnarray}
and the metric tensor do not change
\begin{eqnarray}
g''_{\mu\nu}=\left(\begin{array}{cc}
1 & 0\\ 0 & -1\end{array}\right)
\end{eqnarray}
Here the components already represent physical coordinates
\begin{eqnarray}
X''^{0}=x'^{0}&=&\gamma (x^{0}-\beta x^{1})\\
X''^{1}=x'^{1}&=&\gamma (x^{1}-\beta x^{0})
\end{eqnarray}
where $x^{0}=c t$, $x''^{0}=c t''$, $\beta=\frac{v}{c}$ and
$\frac{1}{\gamma^{2}}=1-\beta^{2}$. It can be written as
\begin{eqnarray}
T''=t''&=&\gamma (t-v x)\\
X''=x''&=&\gamma (x-v t)
\end{eqnarray}
these equations are well-known as Lorentz transformations, with which we most often deal in academic textbooks.

The speed of light in $S'$ now is given by
\begin{eqnarray}\label{vnc}
\Delta s^{2}=d\vec{R}\,d\vec{R}=g'^{\ast}_{\mu\nu}dX'^{\mu}dX'^{\nu}=0\\
V'^{\ast}=\frac{dX'}{dT'}=\pm c \nonumber
\end{eqnarray}
However the de-synchronized transformation lead directly to the physical components.

\section{Discussion and Conclusion}

In this work we unify the Lorentz Transformation and the Inertial transformation in a mathematically consistent way by using Clifford Algebra. The following figure show the relation between the three systems, the system $S$ or the ``rest system'', the de-synchronized transformation (Einstein theory) and Synchronized transformation (Other Theory).

\begin{figure}[H] 
 \centering
 \includegraphics[scale=0.5]{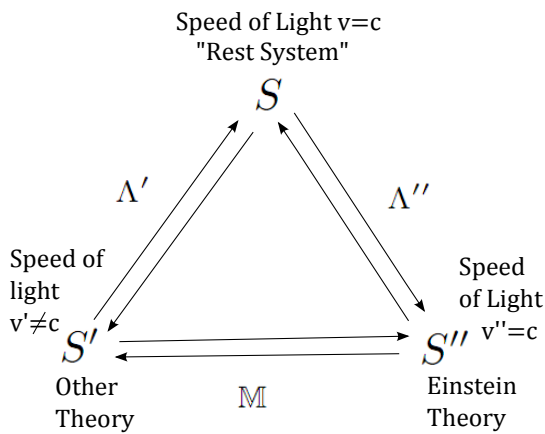}
 \caption{Lorentz Transformation and the Inertial transformation and relation between them.}\label{isometria}
 \end{figure}

According to clifford algebra and differential geometry we find the physical and coordinates components of both transformations. In the case of synchronous transformation, the physical components are exactly the Galilean transformations. But the physical one-way speed of light in $S'$ is not $c$, as shown in the equation-(\ref{vnc}). It is important to note that the transformation shown in equation-(\ref{vnc}) is not the same the old Galilean transformation, because the context is completely different. That's just a coincidence, the coordinates that we are using here belong to bases in a Clifford algebra. Another interesting result is due to the fact the synchronized transformations lead directly to physical components and the physical one-way speed of light in $S'$ as $c$.

We show that both scenarios, de-synchronization Einstein theory and synchronized theory, are all mathematically equivalent through a simpler and more direct way, that is, by means of the Clifford algebra transformation.

\end{document}